\documentclass{ws-procs975x65}
\usepackage{amsmath}
\begin{document}

\title{THE MATTER POWER SPECTRUM AS A TOOL TO DISCRIMINATE DARK MATTER-DARK ENERGY INTERACTIONS}
\author{Germ\'{a}n Olivares$^*$ and Diego Pav\'{o}n$^\dag$}
\address{Departmento de F\'{\i}sica,\\
 Universidad Aut\'{o}noma de Barcelona, \\
 08193 Bellaterra (Barcelona),\\
Spain\\
$^*$E-mail: german.olivares@uab.es\\
$^\dag$E-mail: diego.pavon@uab.es}
\author{Fernando Atrio-Barandela$^\ddag$}
\address{Department of Physics and Astronomy,\\
University of Pennsylvania,\\
Philadelphia, Pennsylvania 19104,\\
USA\\
$^\ddag$E-mail: atrio@usal.es}

\begin{abstract}
The coincidence problem of late cosmic acceleration gets
significantly alleviated when a suitable interaction between
matter and dark energy, either of phantom type or not, enters the
picture. We show that a class of models featuring this interaction
fares rather well when contrasted with the anisotropies of the
CMBR and the matter power spectrum. The latter test is very
sensible to the interaction and may be used to discriminate
between different models.

\end{abstract}

\bodymatter

\section{Introduction}
In order to account for the current stage of cosmic accelerated
expansion many dark energy candidates have been invoked up to now.
By far, the simplest one is the cosmological constant. However,
while models based on it seem to fare rather well when confronted
with the observational data they face two severe difficulties on
the theoretical side: $(i)$ its measured value is about $120$
orders of magnitude lower than the one predicted by QFT, and
$(ii)$ the ``coincidence problem", namely, {\it why are the
densities of matter and dark energy of precisely the same order
today.} This is why dynamic dark energy models, not based on the
zero-point energy of vacuum, have been proposed in the recent
years.

There is a particular class of models in which the dark energy
interacts with the dark matter (assumed non-relativistic) so that
neither of them conserve separately \cite{luca}. Here, we consider
a specific model within this class with the interesting feature
that it evades the first difficulty (the dark energy no longer
relies on the quantum vacuum)  and considerable alleviates the
second one\cite{ladw}. In this model the balance equations for the
cold dark matter and dark energy are $\dot{\rho}_{c} + 3 H\rho_{c}
= Q$, and
$\dot{\rho}_{x}+3H(1+w_{x})\rho_{x} = -Q$, respectively, with $Q =%
3H c^{2} (\rho_{c} +\rho_{x})$, where $c^{2}$ is a dimensionless,
constant parameter that measures the strength of the interaction.
Thus, the ratio, $\rho_{c}/\rho_{x}$, between the aforesaid energy
densities is found to evolve from a constant, but unstable value,
at early times to a lower, constant and stable value at late times
\cite{ladw} -see also Fig. 2 of \cite{gmo1}.

The target of short Communication is  to use the cosmic microwave
background radiation (CMBR) and large scale structure to constrain
this model (for full accounts, see Refs. \cite{gmo1} and
\cite{gmo2}, respectively). As we shall see, the latter can be
used as a tool to discriminate between interacting and
non-interacting models as well as to discriminate between
different interactions. The SNIa data practically do not constrain
the model as the $c^{2}$ shows large degeneracy -see Fig. 3 of
\cite{gmo1}.

\section{Constraints from the CMBR}
Upon exploring the whole parameter space and using the the first
year data provided by the WMAP satellite we found that the
parameters of the model fall into the following ranges
\cite{gmo1}: $\Omega_{x} = 0.43 \pm 0.12$, $\Omega_{b} = 0.08\pm
0.01$, $n_{s} = 0.98\pm 0.02$, $H_{0} = 56 \pm 4$km/s/Mpc. As for
the equation of state parameter of the dark energy only an upper
bound can be set, $ w_{x}\leq -0.86$, while the preferred value of
$c^{2}$ is $5 \times 10^{-3}$ but it still exhibits degeneration.
The Bayesian Information Criteria \cite{bic} allows to show that
this extra parameter fits the data better than models with no
interaction. After combining the WMAP and the SNIa data of Riess
{\it et al.}, $\Omega_{x}$ goes up to $0.68$ and $c^{2}$ increases
to $6.3 \times 10^{-3}$.

\section{Constraints from the matter power spectrum}
For the sake of comparison, we assume that in interacting and
non-interacting models density perturbations have the same
amplitude when they come within the horizon. For non-interacting
models, this prescription leads to the Harrison-Zeldovich power
spectrum, $P(k) \sim k^{-n}$ with $n_{s} = 1$ on large scales.
During the matter epoch, if density perturbations and the
background energy density evolve as $\delta_{c} \sim a^{p/2}$ and
$\rho_{c} \sim a^{-\alpha}$, respectively, we have that $P(k) \sim%
k^{-3+2p/(\alpha-2)}$. During the matter-dominated era, $p \simeq
2(\alpha-2)+0.6\,c^{2}$ and the slope of the scale-invariant
spectrum is $n_{s} = 1$ with a very weak dependence on the
interaction.

The slope of the matter power spectrum on scales $k \geq k_{eq}$
is determined by the growth rate of subhorizon sized matter
perturbations during radiation domination. If a mode that crosses
the horizon before matter radiation equality grows as $\delta_{c}
\sim \tau^{q/2}$ during the radiation dominated era, then the
amplitude of the power spectrum today would be $P(k) =
P(k_{eq})(k_{eq}/k)^{-3+q}$. For cold dark matter models, dark
matter perturbations experience only logarithm growth, so models
with less growth will have less power at small scales as do, for
instance, mixed dark matter models \cite{mixed} which contain a
significant fraction of massive neutrinos. Figs. 2(a) and 2(b) of
Ref. \cite{gmo2} depicts the power spectrum for different
interacting models and mixed dark matter models, respectively.
With increasing $c^{2}$ or $m_{\nu}$ the matter power spectrum
exhibits larger oscillations as a consequence to the increased
ratio of baryons to dark matter. Figs. 2(c) and 2(d), of the same
reference, show that the  slope of $P(k)$ decreases with
increasing $c^{2}$ and $m_{\nu}$. In both cases the behavior is
rather similar. Therefore, observations of large scale structure
that constrain the neutrino mass also serve to set constraints on
the strength of the dark matter-dark energy interaction during the
radiation-dominated era and discriminate interacting from
non-interacting models. Further, there is an important difference
between interacting and mixed models. In the former the maximum of
$P(k)$ shifts to the left for increasing $c^{2}$, in the latter
the maximum does not shifts by increasing the neutrino mass. This
can be understood as follows. At larger $c^{2}$, the dark matter
density becomes smaller at any given redshift and the
matter-radiation equality is delayed. This does not occur with
massive neutrinos where the matter-radiation equality takes place
always at the same redshift.

Since the interaction affects the slope of $P(k)$, we resorted to
the 2dFRGS data \cite{2df} to constrain $c^{2}$. We used a Monte
Carlo Markov chain to run the CMBFAST code, adapted to solve the
interacting model. We ran the chain for $10^{5}$ models, that were
sufficient to reach convergence. Fig. 4 of Ref. \cite{gmo2}
depicts the joint confidence contours at the 68\%, 95\%, and 99\%
level for pairs of parameters after marginalizing over the rest.
The $1\sigma$ confidence levels and upper limits for the model
parameters resulted to be: $c^{2}\leq 3 \times 10^{-3}$,
$\Omega_{c}h^{2} =0.1\pm 0.02$, $H_{0} = 83^{+6}_{-10}$km/s/Mpc.
The data are very insensitive to $w_{x}$ and baryon fraction. It
is only fair to say that quintessence non-interacting models are
also compatible with the 2dFGRS data at $1\sigma$ level.

\section{Concluding remarks}
In summary, $(i)$ the interacting dark matter-dark energy model of
Ref. \cite{ladw} significantly alleviates the coincidence problem
and is consistent with the observational data, $c^{2} < 10^{-2}$
at 99\% confidence level. $(ii)$ $k_{eq}$ is sensitive to the
$c^{2}$ value and decreases with increasing $c^{2}$. $(iii)$ The
slope of the matter power spectrum is also affected by the
interaction -the stronger the interaction, the more negative the
slope. This may be of help to determine $c^{2}$.

\bibliographystyle{ws-procs975x65}

\end{document}